\documentclass[preprint,12pt]{elsarticle}

\usepackage{amssymb}
\usepackage{amsmath,amssymb,amsfonts,latexsym}
\usepackage{CJK}
\usepackage{bm}
\usepackage{indentfirst}
\usepackage{cancel}
\usepackage{float}

\journal{Physics Letter B}

\begin{document}

\begin{frontmatter}



\title{The Emergence of Hopf Term and The Marginality of Dirac model}


\author{Y. X. Zhao}
\author{Z. D. Wang}

\address{Department of Physics and Center of Theoretical and Computational Physics, The University of Hong Kong, Pokfulam Road, Hong Kong}

\begin{abstract}
The subtle relation of the marginality of Dirac model and the emergence of the Hopf term through a Yukawa-type interaction is revealed in this work. We show that the improvement of the marginality of Dirac mode through an infinitesimal non-relativistic term, which is irrelevant for renormalization group, is necessary to emerge the Hopf term. It is found that the appearance of the Hopf term is always accompanied with chiral boundary modes, and the sign of the coupling constant of the infinitesimal non-relativistic term decides their chirality. A lattice Dirac model is also constructed to realize the Hopf term.
\end{abstract}

\begin{keyword}

Quantum anomaly\sep Marginality of Dirac model\sep Hopf term\sep Lattice model


\end{keyword}

\end{frontmatter}

In spatial dimension $d$, the unit vector field $n(x)\in S^d$, which may be resulted from symmetry breaking, can have soliton excitations due to the homotopy group $\pi_d(S^d)\cong\mathbf{Z}$. For $d>2$, these solitons can only have fermionic or bosonic statistics because of $\pi_{d+1}(S^d)\cong\mathbf{Z}_2$, while in two spatial dimensions they may have anyonic statistics since $\pi_{3}(S^2)\cong\mathbf{Z}$\cite{Statistic-Hopf,Dirac-Hopf-4}.  The value of the Hopf term is the corresponding homotopy number for a specific $\mathbf{n}(x)$ in space-time. Thus its coupling constant decides the statistics of the solitons, which are called skyrmions in two spatial dimension. Lots of previous works in both relativistic quantum field theory and condensed matter physics have confirmed that the Hopf term can be emerged out by the Yukawa-type interaction between $\mathbf{n}(x)$ field and fermions, but with coupling constants only for fermionic or bosonic statistics\cite{Dirac-Hopf-4, Volovik-He3,Yakov,Volovik-book,Dirac-Hopf-1,Dirac-Hopf-2,Dirac-Hopf-3,Dirac-Hopf-5}. 

In the present work, we focus on the emergence of the Hopf term from the Yukawa-type interaction with Dirac fermions. Although it is proposed that the Hopf term can be emerged out in this way\cite{Dirac-Hopf-4, Dirac-Hopf-1,Dirac-Hopf-2,Dirac-Hopf-3,Dirac-Hopf-5},
there is a subtleness in the model related to the regularization scheme
and irrelevant terms, which can be exactly summed up as the marginality of the Dirac model from our analysis. The emergence of the Hopf term in this relativistic mode is due to
a homotopy number in the $(\omega,\mathbf{k})$-space just like that
in models in condensed matter physics\cite{Volovik-He3,Yakov,Nonlinear-Hopf,TI-1,TI-2,Xiao-Liang-PRB}. We show that the marginal characteristic of the Dirac model and its improvement
play significant roles in the emergence of the Hopf term. Specifically, adding an infinitesimal non-relativistic term, which is irrelevant for renormalization group, is necessary to emerge the Hopf term. Through
the connection of this model with the theory of topological insulators\cite{TI-1,TI-2,Xiao-Liang-PRB},
we can see that the appearance of the Hopf term is alway accompanied
with chiral boundary excitations. Moreover the irrelevant and infinitesimal term has
physical effects that the sign of its coupling constant can be reflected by the chirality of boundary
modes.

We start with the Dirac Hamiltonian in (2+1) dimensions,
\[
\mathcal{H}(\mathbf{k})=-\gamma_{0}\mathbf{\gamma}\cdot\mathbf{k}+\gamma_{0}m,
\]
where we choose $\gamma_{0}=\tau_{3}$, $\gamma_{1}=i\tau_{1}$ and
$\gamma_{2}=i\tau_{2}$ with $\tau_{i}$ being Pauli matrices, and
alternatively, 
\[
\mathcal{H}(\mathbf{k})=\tau_{2}k_{x}-\tau_{1}k_{y}+\tau_{3}m.
\]
The corresponding Green's function, $G(t,\mathbf{r})=\langle\psi^{\dagger}(t,\mathbf{r})\psi(0,0)\rangle$,
can be expressed in $(\omega,\mathbf{k})$-space and Euclidean signature
as 
\[
G^{-1}(\omega,\mathbf{k})=i\omega-\mathcal{H}(\mathbf{k}).
\]
To introduce the $(\omega,\mathbf{k})$-space topology, we regard
$G^{-1}(\omega,\mathbf{k})$ as a mapping: 
\[
\begin{array}{cccc}
G^{-1}: & \mathcal{M} & \rightarrow & GL(N,\mathbf{C})\\
 & (\omega,\mathbf{k}) & \mapsto & G^{-1}(\omega,\mathbf{k})
\end{array}
\]
with $N=2$ for Dirac Hamiltonian and $\mathcal{M}$ being the manifold
of $(\omega,\mathbf{k})$-space, and then compactify $\mathcal{M}$
as $S^{3}$ by treating the infinity as one point. The smooth mappings
from $S^{3}$ to $GL(N,\mathbf{C})$ can be classified by the homotopy
group $\pi_{3}(GL(N,\mathbf{C}))\cong\mathbf{Z}$. We can judge which
class the $G^{-1}(\omega,\mathbf{k})$ belongs to through computing the
homotopy number by the following formula,

\begin{equation}
N[G]=\frac{1}{24\pi^{2}}\mathbf{tr}\int d\omega d^{2}k\:\epsilon^{\mu\nu\rho}G\partial_{\mu}G^{-1}G\partial_{\nu}G^{-1}G\partial_{\rho}G^{-1},\label{eq:Homotopy number}
\end{equation}
which gives out the homotopy number. However, the infinity is singular
for the Dirac model, because $\mathcal{H}$ approaches different matrices
when $\mathbf{k}$ is going to the infinity along different directions.
The marginal characteristic of the Dirac model lies in the fact that,
if we keep plugging $G^{-1}=i\omega-\mathcal{H}$ into Eq.(\ref{eq:Homotopy number}),
the result is 

\begin{equation}
N=-\frac{1}{2}\mathbf{sgn}(m)\label{eq:Dirac-Chern-Number}
\end{equation}
with $\mathbf{sgn}(a)$ being the sign of $a$, which is a half integer.
To cure this illness at infinity, we modify the Dirac model a little by adding
a term that may be sent to zero after all calculations. We may add a term  preserving the Lorentz invariance, such as $Bk^{2n}$ with $n$ being a positive integer, for instance,
\[
\mathcal{H}_{m}(\mathbf{k})=\tau_{2}k_{x}-\tau_{1}k_{y}+\tau_{3}(m-Bk^{2}).
\]
As for models like this, although the behavior of the Green's function at infinity is cured, we note that the homotopy number calculated by Eq.(\ref{eq:Homotopy number})  is zero. To obtain a nonzero homotopy number, we have to add a non-relativistic term, and for instance modify the Dirac model as
\begin{equation}
\mathcal{H}_{M}(\mathbf{k})=\tau_{2}k_{x}-\tau_{1}k_{y}+\tau_{3}(m-B\mathbf{k}^{2}).\label{eq:MDirac}
\end{equation}
We may send $B$ to zero after all the calculations.  Substituting $G_{M}^{-1}=i\omega-\mathcal{H}_{M}$ into Eq.(\ref{eq:Homotopy number}),
we can obtain
\begin{equation}
N=-\frac{1}{2}(\mathbf{sgn}(m)+\mathbf{sgn}(B)).\label{eq:MDirac-Chern-Number}
\end{equation}
Actually the above result holds for any term with the form $B\mathbf{k}^{2n}$. For a given $m$, the crossing over $B=0$ may send a topologically trivial Green's function to a nontrivial one, or vise versa. 

\begin{figure}
\begin{center}
\includegraphics[scale=0.5]{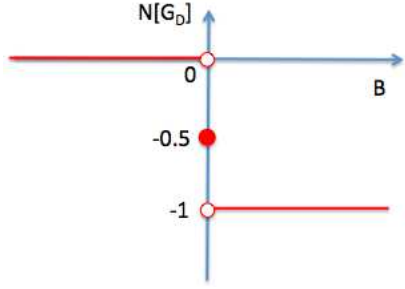}
\end{center}
\caption{The marginal characteristic of the Dirac model\label{CN-of-MDirac}}
\end{figure}

We see that the Dirac model does have the marginality in the sense that an infinitesimal modification of $B$ can send it to a topologically
nontrivial one or a trivial one depending on the sign of $B$, which is illustrated in Fig(\ref{CN-of-MDirac}).  In our case, the homotopy number has geometric meanings. If we look at the whole parameter space of the Green's function, $i.e.,$ the space spanned by $(\omega,\mathbf{k}; B, m)$, there are some points where $G^-1(\omega,\mathbf{k})$ is singular. The homotopy number can be nonzero only if $(\omega,\mathbf{k})$-space as a three-dimensional surface enclose some of these singular points.


Lots of previous works \cite{Dirac-Hopf-1,Dirac-Hopf-2,Dirac-Hopf-3,Dirac-Hopf-4,Dirac-Hopf-5}
reveal that Hopf term can be emerged out by coupling the Dirac model
to a unit-vector field $\mathbf{n}(x)$ (which may be an $O(3)$-order parameter after symmetry breaking) by a Yukawa-type interaction, $i.e.$,
\begin{equation}
\mathcal{L}=\bar{\psi}\left(i\gamma^{\mu}\partial_{\mu}-m\mathbf{n}\cdot\mathbf{\sigma}\right)\psi,\label{eq:Dirac-O(3)}
\end{equation}
where $\mathbf{n}(x)\in S^2$ varies slowly in space-time, and $\sigma_{i}$
are pauli matrices acting on the isospin space of $\psi$. However there exists
a subtleness related to the marginal character of the Dirac model,
which is not manifestly presented in the previous works. One of main
purposes of this work is to elucidate this subtleness. The key trick
to deduce the Hopf term is to express $\mathbf{n}(x)\cdot\sigma$
as \cite{Volovik-He3,Dirac-Hopf-2,Nonlinear-Hopf}
\[
\mathbf{n}(x)\cdot\mathbf{\sigma}=U^{\dagger}(x)\sigma_{3}U(x)
\]
 with $U(x)\in SU(2)$, and introduce the auxiliary gauge field
\[
\Omega_{\mu}=iU\partial_{\mu}U^{\dagger}.
\]
The Hopf term can be expressed in terms of the auxiliary gauge field $a_\mu=\langle \mathbf{n}|i\partial_\mu|\mathbf{n}\rangle$\cite{Dirac-Hopf-2,Statistic-Hopf}, where $|\mathbf{n}\rangle$ being the eigenstate of $\mathbf{n}\cdot\sigma$ with eigenvalue $+1$, as 
\[
H_{Hopf}=-\frac{1}{4\pi^{2}}\int d^{3}x\:\epsilon^{\mu\nu\lambda}a_{\mu}\partial_{\nu}a_{\lambda}.
\]
Since $|\mathbf{n}\rangle=U^{\dagger}|\uparrow\rangle$, The Hopf term can also be expressed in terms of $\Omega_\mu=\Omega^i_\mu\sigma^i$ as
\[
H_{Hopf}=-\frac{1}{16\pi^{2}}\int d^{3}x\:\epsilon^{\mu\nu\lambda}\Omega_{\mu}^{3}\partial_{\nu}\Omega_{\lambda}^{3}.
\]

 After making the gauge transformation $\psi\rightarrow U\psi$ and
integrating over the fermionic freedoms, we can obtain the Hopf term by dropping higher order terms in the gradient expansion, 
\begin{eqnarray}
\mathcal{L}_{Hopf} & = & -\frac{i}{16\pi}N\int d^{3}x\:\epsilon_{\alpha\beta\gamma}\Omega_{3}^{\alpha}\partial^{\beta}\Omega_{3}^{\gamma}(x)\nonumber \\
 & = & i\pi N[G]H_{Hopf}\label{eq:L_Hopf}
\end{eqnarray}
where $N$ is given by Eq.(\ref{eq:Homotopy number}) and in this
case 
\[
G^{-1}=\begin{pmatrix}i\omega-\mathcal{H}(\mathbf{k};m)\\
 & i\omega-\mathcal{H}(\mathbf{k};-m)
\end{pmatrix}.
\]
In the language of Feynman diagrams, this term comes from two-vertex and three-vertex diagrams. 

Since $G^{-1}$ is diagonal, using Eq.(\ref{eq:Dirac-Chern-Number}) we have 
\[
N[G]=N[G^{00}]+N[G^{11}]=-\frac{1}{2}(\mathbf{sgn}(m)+\mathbf{sgn}(-m))=0.
\]
Thus exactly speaking, there is no Hopf term emergent from Dirac model
without certain terms to improve the marginal characteristic of the
Dirac model. Obviously we can use Eq.(\ref{eq:MDirac}) to do the improvement,
however the appearance of Hopf term depends on the way to add the
term of $B\mathbf{k}^{2}$. If we simply substitute $m$ with $m-B\mathbf{k}^{2}$ in
Eq.(\ref{eq:Dirac-O(3)}), it is easy to confirm that the homotopy
number is still zero using Eq.(\ref{eq:MDirac-Chern-Number}). The
right way is to modify Eq.(\ref{eq:Dirac-O(3)}) as 
\begin{equation}
\mathcal{L}_{M}=\bar{\psi}\left(i\gamma^{\mu}\partial_{\mu}-m\mathbf{n}\cdot\mathbf{\sigma}-B\nabla^{2}\right)\psi,\label{eq:MDirac-O(3)}
\end{equation}
which corresponds to 
\[
G_{M}^{-1}=\begin{pmatrix}i\omega-\mathcal{H}_{M}(\mathbf{k};m,B)\\
 & i\omega-\mathcal{H}_{M}(\mathbf{k};-m,B)
\end{pmatrix}
\]
with $\mathcal{H}$ given by Eq.(\ref{eq:MDirac}). From Eq.(\ref{eq:MDirac-Chern-Number}),
we obtain 
\[
N[G_{M}]=\mathbf{sgn}(B).
\]
Thus an infinitesimal additional term of $B\mathbf{k}^{2}$ can ensure the
emergence of the Hopf term in Eq.(\ref{eq:Dirac-O(3)}). This term is irrelevant in
the sense of renormalization group, but its sign does have physical
meanings when we consider the boundary of the system. In Eq.(\ref{eq:Dirac-O(3)})
or Eq.(\ref{eq:MDirac-O(3)}), we assume that $\mathbf{n}(x)$ varies
slowly enough in spacetime. So when treating $\mathbf{n}(x)$ as a
constant, we see $N[G]$ is just the Chern-number of the corresponding
fermionic system as a topological insulator. The nontrivial Chern-number
$N$ implies there are $N$ flavors of chiral fermions on the boundary,
and the sign of $N$ indicates the chirality of the boundary excitations\cite{Volovik-book,IndexTheorem,TI-1,TI-2}.
As for Eq.(\ref{eq:MDirac-O(3)}), if the irrelevant term has positive(negative)
coupling constant $B$, the boundary excitations are left-handed(right-handed)
fermions. Thus though different signs of $N$ in Eq.(\ref{eq:L_Hopf})
implies the same statistics of the skyrmions, but they imply different
chiralities of boundary excitations. Moreover, as the Hopf term is
always accompanied with the $(\omega,\mathbf{k})$-space homotopy
number in Eq.(\ref{eq:L_Hopf}), the fermionic statistics of skyrmions
indicates the chiral boundary excitations.

We carry on to the lattice model. The square lattice version of Dirac model is 
\begin{eqnarray*}
H_{0} & = & \sum_{\vec{r},i}\left[\psi_{\vec{r}}^{\dagger}\left(\frac{c\tau_{3}-i\tau_{i}}{2}\right)\psi_{\vec{r}+\hat{i}}+h.c.\right]+M\sum_{\vec{r}}\psi_{\vec{r}}^{\dagger}\tau_{3}\psi_{\vec{r}}\\
 & = & \sum_{k}\psi_{k}^{\dagger}\left[\sum_{i}\sin k_{i}\gamma^{i}+\left(M+c\sum_{i}\cos k_{i}\right)\gamma^{0}\right]\psi_{k},
\end{eqnarray*}
where the two sublattices have different on-site energy, $i.e.$,
one has $M$ and the other has $-M$. It describes a topological insulator
with the topological number calculated by Eq.(\ref{eq:Homotopy number})\cite{Xiao-Liang-PRB},
that is 
\begin{equation}
N=\begin{cases}
1 & 0<M<2c\\
-1 & -2c<M<0\\
0 & otherwise
\end{cases}.\label{eq:Lattice-Chern-Number}
\end{equation}
Without losing any generality, we have assumed $c>0$. To obtain the Hopf term from this lattice fermionic model, we add an appropriate coupling term to the Hamiltonian:
\[
H_{1}[\mathbf{n}]=\lambda\mathbf{n}_{\vec{r}}\cdot\mathbf{\sigma}
\]
with $\lambda>0$, which may describe a spin interaction. Then the
homotopy number as the coupling constant of the Hopf term is calculated by the following Green's function.
\[
G_{L}^{-1}(\omega,\mathbf{k})=\begin{pmatrix}G_{\uparrow\uparrow}^{-1}(\omega,\mathbf{k})\\
 & G_{\downarrow\downarrow}^{-1}(\omega,\mathbf{k})
\end{pmatrix},
\]
where
\begin{eqnarray*}
G_{\uparrow\uparrow}^{-1}(\omega,\mathbf{k}) & = & i\omega-\left(\sin k_{x}\tau_{x}+\sin k_{y}\tau_{y}\right)\\
 &  & -\left(M+\lambda+c\sum_{i}\cos k_{i}\right)\tau_{z}
\end{eqnarray*}
and
\begin{eqnarray*}
G_{\downarrow\downarrow}^{-1}(\omega,\mathbf{k}) & = & i\omega-\left(\sin k_{x}\tau_{x}+\sin k_{y}\tau_{y}\right)\\
 &  & -\left(M-\lambda+c\sum_{i}\cos k_{i}\right)\tau_{z}
\end{eqnarray*}

\begin{figure}
\begin{center}
\includegraphics[scale=0.3]{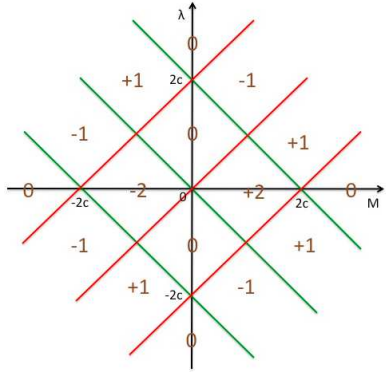}
\end{center}
\caption{The homotopy number diagram. The black lines are coordinates of the $M-\lambda$ plane. Green and red lines separate the plane into several regions. Each region has its own homotopy number that is written in brown. \label{Phase-Diagram}}
\end{figure}

Using Eq.(\ref{eq:Lattice-Chern-Number}), we can calculate the homotopy number for a specific pair of $M$ and $\lambda$. The result is shown in Fig.[\ref{Phase-Diagram}]. For regions with homotopy number $\pm1$, the skyrmions are fermions, while for the others with $\pm2$ or $0$, the skyrmions are bosons. However, given $\mathbf{n}(x)$ varies slowly enough, a nonzero homotopy number implies chiral boundary modes with chirality being decided by its sign. 

To conclude, we illustrated the marginality of the Dirac model, and revealed its relation to the emergence of the Hopf term. We showed that, to emerge the Hopf term from Dirac model, it is necessary to improve the marginality through adding an infinitesimal non-relativistic term, which has the physical consequence that its sign decides the chirality of boundary modes. We also constructed a lattice Dirac model to emerge the Hopf term.

This work was supported by the GRF (HKU7058/11P) and the CRF (HKU8/11G) of Hong Kong.

\bibliographystyle{model1-num-names}
\bibliography{<your-bib-database>}

\end{document}